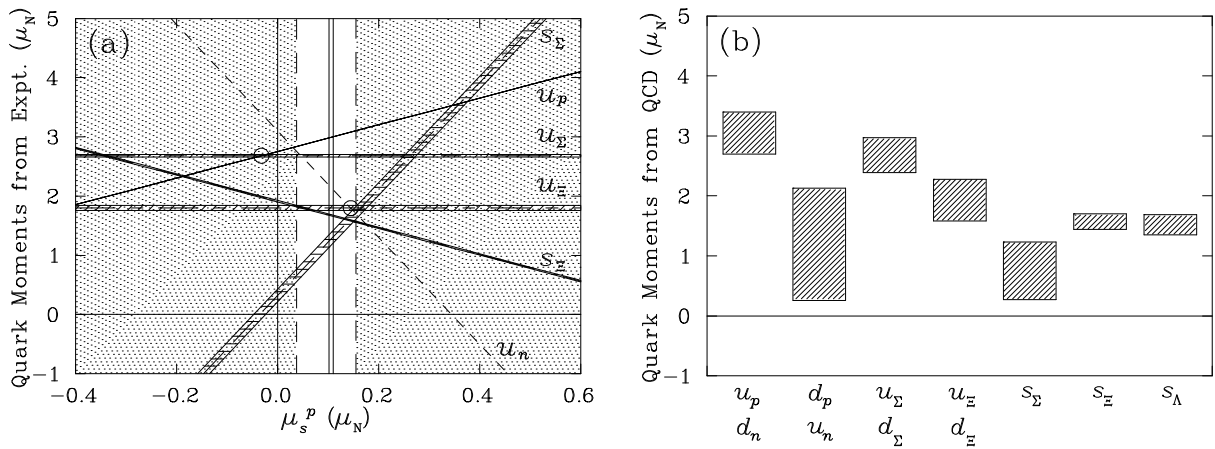

Figure 1. Effective quark moments extracted from experiment as a function of $\mu_s^p$ (a) are compared with the best estimates available from QCD (b).

## 2. DISCUSSION

To determine the preferred values for $\mu_s^p$ we turn to the best estimates of effective quark moments available from *ab initio* QCD [1]. The raw lattice QCD results must be corrected for order $a$ lattice spacing errors in the standard Wilson fermion action, and finite volume errors. A single scale parameter is introduced to account for the leading order effects of these systematic errors. Uncertainties due to the quenched approximation or extrapolations are not expected to be important. Chiral perturbation theory ($\chi$PT) indicates the leading-order nonanalytic corrections are proportional to $m_\pi$; they are negligible in the regime where $\chi$PT may be regarded as reliable. The scale parameter is fixed by matching the lattice and experimentally determined value for the effective quark moment of the $u$ quark in $\Sigma$, which is independent of $\mu_s^p$, $u_\Sigma = (\Sigma^+ - \Sigma^-)/2$.

Figure 1a illustrates the effective quark moments extracted from (1) using experimental baryon moments. The quark moments, normalized to unit charge and plotted as a function of $\mu_s^p$, are compared with those determined from QCD in 1b. We note that a value for $\mu_s^p$ which reproduces the effective moments of the simple quark model does not exist. Even the conservative assumptions of Ref. [5], $u_p = u_\Sigma$ and $u_n = u_\Xi$, marked by o in 1a, are mutually exclusive points. The shaded regions mark the values of $\mu_s^p$ in which at least one effective moment deviates from the QCD prediction by more than $2\,\sigma$. Strangeness in the nucleon is essential to maintain agreement for both $u_n$ and $s_\Xi$ within $2\,\sigma$. The double line at $\mu_s^p \simeq 0.11\ \mu_N$ marks the region where all effective moments agree within $1\sigma$.

# Essential strangeness in nucleon magnetic moments

Derek B. Leinweber[a][*]

[a] Department of Physics, The Ohio State University
174 West 18th Avenue, Columbus, OH 43210-1106, USA

Effective quark magnetic moments are extracted from experimental measurements as a function of the strangeness magnetic moment of the nucleon. Assumptions made in even the most general quark model analyses are ruled out by this investigation. *Ab initio* QCD calculations demand a non-trivial role for strange quarks in the nucleon. The effective moments from QCD calculations are reproduced for a strangeness magnetic moment contribution to the proton of 0.11 $\mu_N$, which corresponds to $F_2^s(0) = -0.33\ \mu_N$.

## 1. INTRODUCTION

In this investigation, we take an unusual approach and extract the effective magnetic moments of quarks bound in baryons from the experimentally measured moments. The effective moments are defined as in the QCD investigations of [1,2] via SU(6) wave functions, which get the bulk of the phenomenology right. The focus is then placed on the relevant QCD dynamics essential to reproduce the pattern of moments determined in [2].

There is an additional degree of freedom outside of the standard valence sector of SU(6) which must be taken into account before the effective quark moments may be extracted [3]. This extra degree of freedom, demanded by the Pauli exclusion principle, arises from electromagnetic current interactions with disconnected sea quark loops which in turn interact via gluon exchange. The effective quark moments ($u_p$, $d_p$, $u_\Sigma$, ...) are defined by

$$p \equiv \frac{4}{3}u_p - \frac{1}{3}d_p + \mu_l\,, \quad \Sigma^+ \equiv \frac{4}{3}u_\Sigma - \frac{1}{3}s_\Sigma + \mu_l\,, \quad \Xi^0 \equiv \frac{4}{3}s_\Xi - \frac{1}{3}u_\Xi + \mu_l\,, \qquad (1)$$
$$n \equiv \frac{4}{3}d_n - \frac{1}{3}u_n + \mu_l\,, \quad \Sigma^- \equiv \frac{4}{3}d_\Sigma - \frac{1}{3}s_\Sigma + \mu_l\,, \quad \Xi^- \equiv \frac{4}{3}s_\Xi - \frac{1}{3}d_\Xi + \mu_l\,.$$

Here $p$ represents the magnetic moment of the proton and similarly for the others. $\mu_l$ is the contribution from the disconnected $u, d, s$-quark loops, and is assumed to be equal for all baryons. This approximation is expected to be excellent among octet baryons of a given strangeness [3]. The usual constraint equations, $u_p = -2d_n$, $u_n = -2d_p$, $u_\Sigma = -2d_\Sigma$ and $u_\Xi = -2d_\Xi$, reduce the number of degrees of freedom on the RHS of (1) to seven.

The effective quark moments are defined as a function of the disconnected sea quark contribution, $\mu_l$, by inverting these relations. The relative role of strange and light quarks in the disconnected loop may be estimated through the use of hadronic models [4]. There the contribution of the heavier strange quark is suppressed by a factor of two relative to the $d$ quark. It follows that $\mu_l = -\mu_s^p$, and this relation is used in the following.

[*]This research is supported by the National Science Foundation.